\documentclass[
    aps,
    fleqn,
    a4paper,
    superscriptaddress,
    preprint,
    preprintnumbers,
    showpacs,
    showkeys
]{revtex4}

\usepackage{graphicx}
\usepackage{picins}
\usepackage{amsmath}
\usepackage{amsfonts}
\usepackage[cp1250]{inputenc}

\newcommand{\bE}{\mbox{\boldmath$E$}}
\newcommand{\bP}{\mbox{\boldmath$P$}}
\newcommand{\bD}{\mbox{\boldmath$D$}}

\newcommand{\embox}[1]{{\em #1\/}}

\begin{document}

\preprint{{\em Ferroelectrics\/}: Vol. 238, No. 1-4, Pgs. 767-773 (2000) - {\em (Revised version)}}

\title{
On the extrinsic piezoelectricity
}

\author{A. \surname{Kopal}}
\email{antonin.kopal@tul.cz}
\affiliation{Dept. of Physics, International Center for Piezoelectric Research, Technical University Liberec, Liberec 1, 461 17 Czech Republic}
\author{P. \surname{Mokr\'{y}}}
\email{pavel.mokry@tul.cz}
\affiliation{Dept. of Physics, International Center for Piezoelectric Research, Technical University Liberec, Liberec 1, 461 17 Czech Republic}
\author{J. \surname{Fousek}}
\affiliation{Dept. of Physics, International Center for Piezoelectric Research, Technical University Liberec, Liberec 1, 461 17 Czech Republic}
\affiliation{Materials Research Laboratory, The Pennsylvania State University, State College, PA 16801, USA}
\author{T. \surname{Bahn\'{\i}k}}
\affiliation{Dept. of Physics, International Center for Piezoelectric Research, Technical University Liberec, Liberec 1, 461 17 Czech Republic}

\date{\today}

\begin{abstract}
This work presents a continuation of our last paper, concerning the
theory of the response of an antiparallel domain structure in a
plate-like electroded sample to external electric field. The theory is
based on the exact formula for free energy of the system, formed of a
central ferroelectric part, isolated from electrodes (with a defined
potential difference) by a surface layers. Our calculations are
applicable also to thin films. It is usual to use the term `extrinsic'
for the contribution of domain walls displacement to macroscopic
properties of a sample. In our last paper we discussed the extrinsic
contribution to permittivity. In this work we concentrate on extrinsic
contribution to piezoelectric coefficients in ferroelectrics which are
simultaneously ferroelastics. As an example, we calculate the extrinsic
contribution to $d_{36}$ piezoelectric coefficient of $\rm Rb H_2 P
O_4$, that was recently measured in a wide range of temperature below
Curie point.
\end{abstract}


\maketitle

\section{Introduction}

Samples of ferroelectric single crystals often posses a surface
layers. Its existence greatly influences properties of bulk
samples\cite{ArtA2:102,ArtA2:110,ArtA2:94,ArtA2:342} as well as
of thin films.\cite{ArtA2:255} Equilibrium domain structure in
the system, mentioned above in the abstract, and the role of the
surface layers was first discussed by  Bjorkstam and
Oettel\cite{ArtA2:625} in a special case of shorted electrodes.
In our recent paper\cite{ArtA2:kopal2} we reconsidered this
problem in a general case of nonzero voltage between electrodes,
discussing the response of the domain structure to external
electric field. Our calculations are valid also for thin films and
present, in fact, continuation of our discussion of domain
structures of thin films\cite{ArtA2:kopal1}.

In\cite{ArtA2:kopal2} we used our theoretical results for
prediction of extrinsic contribution to permittivity of the
sample. In the next two sections we give a short review of
notation, description of the model and basic results
from\cite{ArtA2:kopal2}. In last two sections we discuss as an
example the extrinsic contribution to $d_{36}$ piezoelectric
coefficient of $\rm RbH_2PO_4$. We compare our predictions with
the recent measurements of record values $\sim 4000\rm\,pC/N$ in
temperature range 35\,K under critical temperature
146\,K\cite{ArtA2:stula} (see also\cite{ArtA2:nakamura}).

\section{Description of the model}

We consider a plate-like electroded sample of infinite area with
major surfaces perpendicular to the ferroelectric axis $z$.
Central ferroelectric part with antiparallel domains (2.) is
separated from the electrodes (0.), (4.) by nonferroelectric
layers (1.),(3.) (see Fig.\,\ref{artA2:fig:geometry}). The spatial
distribution of the electric field $\bE$ is determined by the
applied potential difference $V=\varphi^{(4)}-\varphi^{(0)}$ and
by the bound charge div$\bP_0$ on the boundary of ferroelectric
material, where $\bP_0$ stands for spontaneous polarization.
Geometrical, electrical and material parameters of the system are
shown in Fig.\,\ref{artA2:fig:geometry}.

We further introduce the symbols
\begin{displaymath}
    c = \sqrt{\varepsilon_a/\varepsilon_c}, \qquad
	 g = \sqrt{\varepsilon_a\,\varepsilon_c},
\end{displaymath}
and several geometrical parameters: the slab factor
\[B=\frac{d}{h},\]
%
%
\parpic[r][t]{
    \begin{minipage}[t]{90mm}
    \begin{minipage}[t]{\textwidth}
        \makebox[\textwidth][t]{
            \includegraphics[width=90mm]{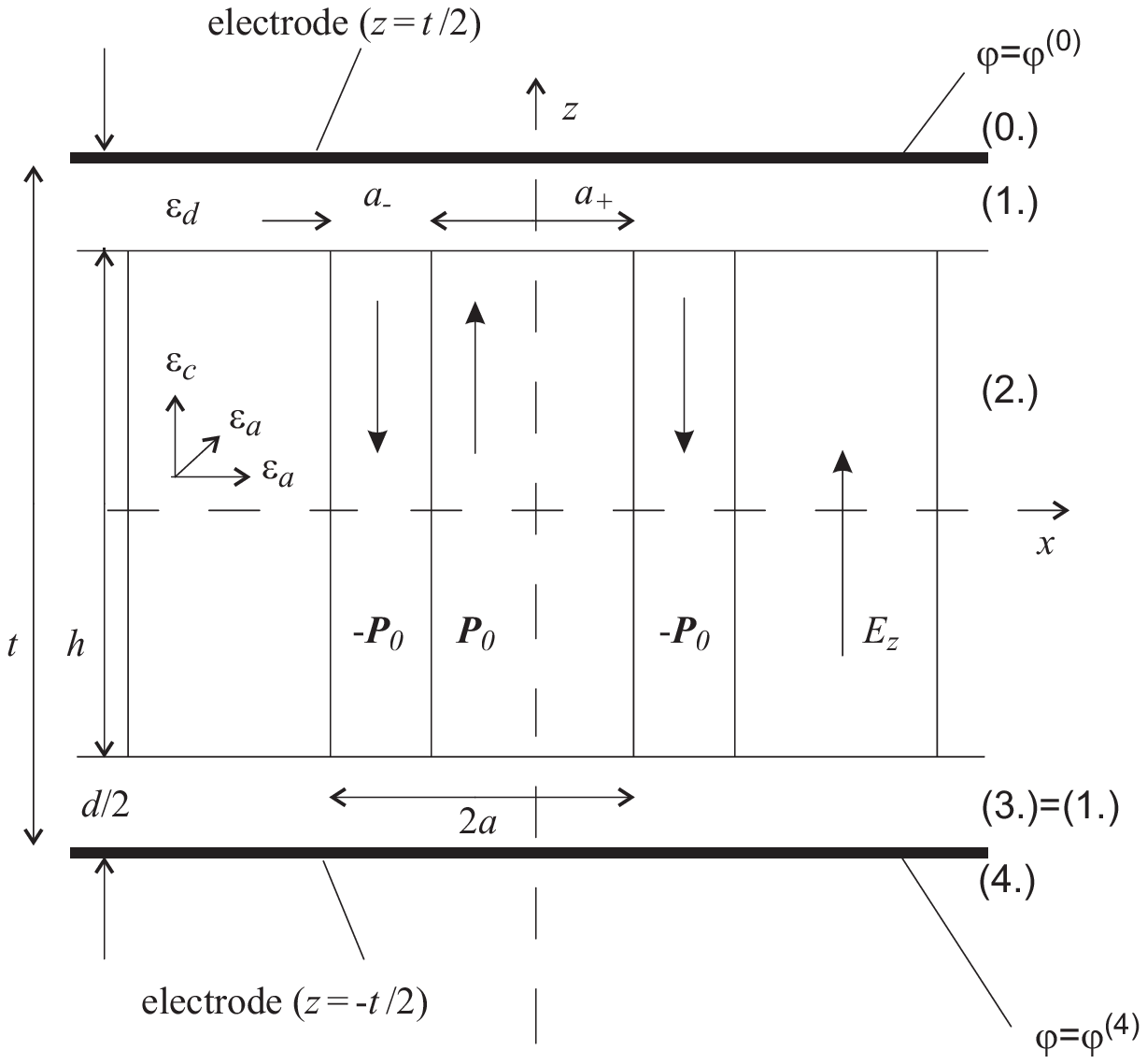}
        }
    \caption{Geometry of the model \label{artA2:fig:geometry}}
    \end{minipage}
    \vspace{5ex}
    \end{minipage}
}
%
the domain pattern factor
\[R = \frac{\pi h}{2a}, \quad 2a = a_+ + a_-, \]
and the asymmetry factor
\[A = \frac{a_+ - a_-}{a_+ + a_-}\ .\]
The ferroelectric material itself is approximated by the equation
of state
\begin{eqnarray}
    D_{x} &=& \varepsilon_0\varepsilon_a E_{x},
    \nonumber \\
    D_z &=& \varepsilon_0\varepsilon_c E_z + P_{0},
    \nonumber
\end{eqnarray}
where $P_{0}$ is the spontaneous polarization along the
ferroelectric axis. This linear approximation limits the validity
of our calculations to the temperature region not very close below
the transition temperature $T_c$. Domain walls are assumed to have
surface energy density $\sigma_w$ and zero thickness.
%

\section{Gibbs electric energy of the system, equilibrium domain structure}

Rather cumbersome calculations\cite{ArtA2:kopal2} lead to the
following formula for Gibbs electric energy per unit area of the
system (in $\rm J\,m^{-2}$), which includes the domain wall
energy, the electrostatic energy whose density is $(1/2)\bE\cdot
(\bD-\bP_0)$ and the work performed by external electric sources
$-VQ$, where $Q$ is the charge on positive electrode. 
\begin{equation}
    G =
        \frac{2}{\pi}\sigma_w R
        +
        P_0 A\
        \frac{
            P_0 A\, B t
            -
            2\varepsilon_0\varepsilon_d V \left( 1 + B \right)
        }{
            2\varepsilon_0
            \left(1 + B\right)
            \left(\varepsilon_d + B\,\varepsilon_c \right)
        }
        +
        \frac{4 P_0^2 t}{\varepsilon_0\pi^2 R (1+B)}
        \sum^{\infty}_{n=1}
        \frac{
            \sin^2{(n\pi(1+A)/2)}
        }{
            n^3(\coth{n B R} +g\coth{n R c})
        }\ .
    \label{artA2:eq:depolenergy}
\end{equation}
The first term represents domain wall contribution while the last one is
the depolarization energy. In the second term we recognize the effect of
layers (1.) and (3.) and of the applied voltage.

In this model we neglect the mechanical interactions between
components of the system. For given slab factor $B$ and voltage
$V$, the equilibrium domain structure, characterized by $R_{\rm
eq}(V)$ and $A_{\rm eq}(V)$, corresponds to local minimum of $G$.
In general a minimum can be found by numerical methods, but for
$BR \gg 1$ and $Rc\gg 1$, the $R_{\rm eq}(V)$ and $A_{\rm eq}(V)$
can be approximated by explicit formulae. For purpose of this
paper we use the following formula for $A_{\rm eq}(V)$
\begin{equation}
    A_{\rm eq}(V) \cong
    \frac{\varepsilon_0 V}{t P_0}
    \left[
        \frac{B}{\varepsilon_d(1+B)} - X
    \right]^{-1}
    \label{artA2:eq:aeqv}
\end{equation}
where
\[
    X =
    \frac{
        2\,\ln{2}
        \left(1+ B\varepsilon_c/\varepsilon_d\right)
    }{
        R^0_{\rm eq}\,(1+B)
        (\varepsilon_d + \sqrt{\varepsilon_a\, \varepsilon_c} )
    }
\]
is considered as a small correction and $R^{0}_{\rm eq}$ is
equilibrium value of $R$ for zero voltage $V$. For the extrinsic
contribution to permittivity of the sample we get from
(\ref{artA2:eq:aeqv}) (see also\cite{ArtA2:kopal2})
\begin{equation}
    \varepsilon_{\rm ext} =
    \left\{
        \left(
            1+B\,
            \frac{
                \varepsilon_c
            }{
                \varepsilon_d
            }
        \right)
        \left[
            \frac{B}{(1+B)\,\varepsilon_d}
             - X
        \right]
    \right\}^{-1}
\label{artA2:eq:eext}
\end{equation}

\section{Extrinsic piezoelectricity}

\begin{figure}[t]
\begin{center}
\begin{minipage}[t]{\textwidth}
\begin{minipage}[t]{0.49\textwidth}
 \makebox[\textwidth][t]{
   \includegraphics[
      width=\textwidth,
      ]{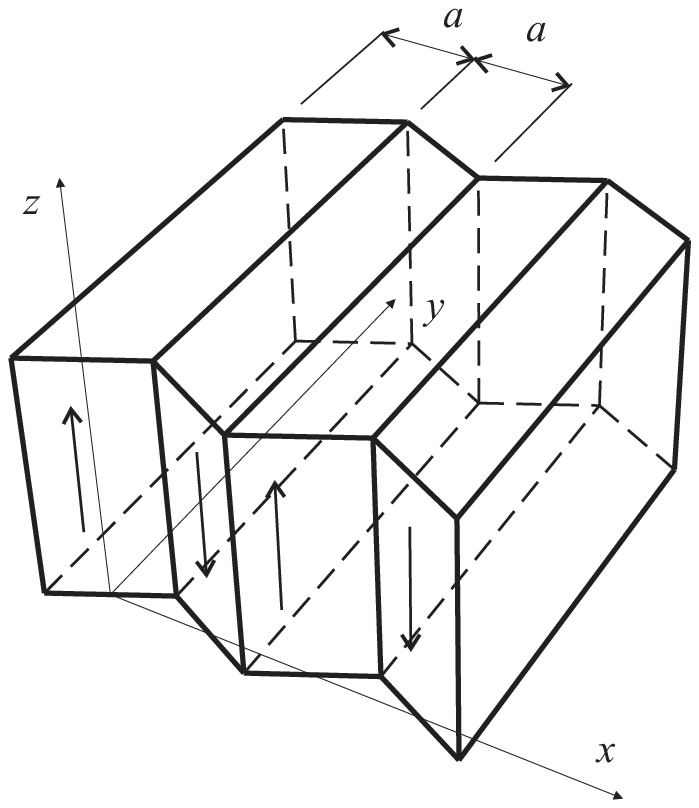}
 }
 \makebox[\textwidth][b]{\hfill a) \hfill}
\end{minipage}
\hfill
\begin{minipage}[t]{0.49\textwidth}
 \makebox[\textwidth][t]{
   \includegraphics[
      width=\textwidth
      ]{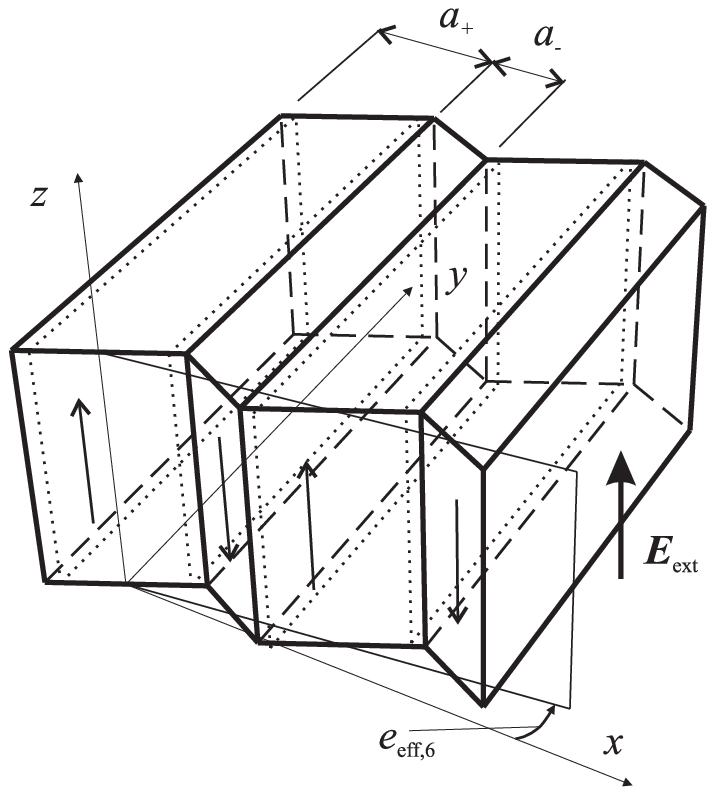}
 }
 \makebox[\textwidth][b]{\hfill b) \hfill}
\end{minipage}
\vspace{3ex}
\end{minipage}
 \begin{minipage}{11cm}
   \caption{Symmetric domain structure with spontaneous strain $e_{0,6}$,
   (a) $E_{\rm ext}=0, A=0$, (b) $E_{\rm ext} \neq 0, A \neq 0$ \label{artA2:fig:symds} \label{artA2:fig:asymds}}
 \end{minipage}
\end{center}
\end{figure}

As an example, in this section we work out the approximate
prediction of extrinsic contribution to $d_{36}$ of $\rm Rb H_2
PO_4$ (RDP), based on our simple model. RDP is a ferroelastic with
spontaneous strain $e_{0,6}$, opposite in opposite polarized
domains. In the Fig.\ref{artA2:fig:symds}a there is $x-y$ cut
through symmetric domain structure ($A=0$, $E_{\rm ext} = V/t =
0$). The situation after application of $E_{\rm ext}$ is shown in
Fig.\ref{artA2:fig:asymds}b ($A \neq 0$). A simple geometric
consideration leads to the formula for average strain $e_{\rm
eff,6}$ of the sample (we neglect the mechanical coupling of the
central part with the rest of the sample)
\begin{equation}
    e_{\rm eff,6} = e_{0,6} A \label{artA2:eq:avgstrain}
\end{equation}

For the extrinsic coefficient $d_{36}$ we get from
(\ref{artA2:eq:aeqv}) and (\ref{artA2:eq:avgstrain})
\begin{equation}
    d_{36} \equiv \frac{e_{\rm eff,6}}{E_3} =
        \frac{\varepsilon_0  e_{0,6}}{P_0}\,
    \left[
        \frac{B}{(1+B)\,\varepsilon_d}
        -
        X
    \right]^{-1} \label{artA2:eq:d36}
\end{equation}
neglecting the small $X$ and for $B\ll 1$ we get more simple
formula\footnote{The limit $B\to 0$ in (\ref{artA2:eq:d36simple})
is not correct, because the assumptions needed for validity of
(\ref{artA2:eq:aeqv}) are violated if $B$ is very small.}
\begin{equation}
    d_{36} = \frac{\varepsilon_0\varepsilon_d e_{0,6}}{P_0B}\ .
    \label{artA2:eq:d36simple}
\end{equation}

\section{Discussion}

An extremely high $d_{36}$ under $T_c$ for RDP was first reported
in\cite{ArtA2:shuvalov}. In a recent
paper\cite{ArtA2:sidorkin}, Sidorkin deduced the dispersion law
of domain wall vibrations, however, in his treatment the existence
of a surface layers is not explicitly considered. We can fit our
theoretical results to measured ones\cite{ArtA2:stula} -
$d_{36,\rm ext}\doteq 4\cdot 10^{-9}\,{\rm C/N}, \varepsilon_{\rm
z,ext} = 2000$ in the 35\,K range plateau under $T_c$. Using
values that roughly apply to RDP: \footnote{$P_0= 4\cdot 10^{-2}
{\rm C/m^2}, \varepsilon_d \doteq \varepsilon_c \doteq 100,
e_{0,6}\doteq 0.015$, see\cite{ArtA2:stula}.} in
(\ref{artA2:eq:d36simple}) resp. (\ref{artA2:eq:eext}) we come to
an agreement for reasonable value of  $B\doteq 0.04$. Naturally
for lower temperatures, the motion of the walls is limited by
``freezing" and both $d_{\rm 36,ext}$ and $\varepsilon_{\rm
z,ext}$ decrease to zero. It is also interesting, for measurements
in\cite{ArtA2:stula} with alternating $E_{\rm ext}\doteq 20\,\rm
V/m$, that corresponding amplitude of alternating $A$ from
(\ref{artA2:eq:aeqv}) is only $10^{-5}$ and displacement of the
walls with $a \doteq 10\,\rm \mu m$ is of the order $10^{-10}\,\rm
m$.

\acknowledgements 

This work has been supported by the Ministry of Education of the Czech
Republic grants CEZ: J11/98:242200002 and VS 96006.


\begin{thebibliography}{99}
\bibitem{ArtA2:102}
R. C. Miller and A. Savage, \embox{J. Appl. Phys.} {\bf 32} (1961).

\bibitem{ArtA2:110}
M. E. Drougard and R. Landauer, \embox{J. Appl..Phys} {\bf 30} (1959).

\bibitem{ArtA2:94}
H. E. M\"{u}ser, W. Kuhn and J. Albers, \embox{Phys. Stat. Sol.(a)}
{\bf 49} (1978).

\bibitem{ArtA2:342}
D. R. Callaby, \embox{J. Appl. Phys.} {\bf 36} (1965).

\bibitem{ArtA2:255}
A. K. Tagantsev, C. Pawlaczyk, K. Brooks and N. Setter,
\embox{Integrated ferroelectrics} {\bf 4} (1994).

\bibitem{ArtA2:625} J. L. Bjorkstam and R. E. Oettel, \embox{Phys. Rev.}
{\bf 159} (1967).

\bibitem{ArtA2:kopal2}
A. Kopal, P. Mokr\'{y}, J. Fousek, T. Bahnik, to appear in
Ferroelectrics.

\bibitem{ArtA2:kopal1}
A. Kopal, T. Bahn\'{\i}k and J. Fousek, \embox{Ferroelectrics}
{\bf 202} (1997).

\bibitem{ArtA2:stula}
M.\,\v{S}tula, J.\,Fousek, H.\,Kabelka, M.\,Fally and H.\,Warhanek,
\embox{J.\,Kor.\,Phys.\,Soc.\,(Proc. Suppl.)} {\bf 32} (1998).

\bibitem{ArtA2:nakamura}
E. Nakamura, \embox{Ferroelectrics} {\bf 135} (1992).

\bibitem{ArtA2:shuvalov}
L. A. Shuvalov, I. S. Zheludev, A. V. Mnatskanyan and Ts.-Zh. Ludupov, I.
Fiala, \embox{Bull. Acad. Sci. USSR, Phys. Ser.} {\bf 31} (1967).

\bibitem{ArtA2:sidorkin}
A. S. Sidorkin, \embox{J. Appl. Phys.} {\bf 83} (1998).

\end{thebibliography}
\end{document}